# On the isotope effects of $ZrX_2$ (X = H, D and T): A First-principles study

D. Chattaraj[a], S.C. Parida[a], Smruti Dash[a], C. Majumder[b,*]

[a]Product Development Division, [b]Chemistry Division

Bhabha Atomic Research Centre, Trombay, Mumbai 400 085, India

ABSTRACT

Zirconium hydride is an important material for storage of hydrogen isotopes. Here we report the structural, electronic, vibrational and thermodynamic properties of $ZrH_2$, $ZrD_2$ and $ZrT_2$ using density functional theory (DFT). The structural optimization was carried out by the plane-wave based pseudo-potential method under the generalized gradient approximation (GGA) scheme. The electronic structure of the $ZrH_2$ compound was illustrated in terms of the electronic density of states, charge density distribution and the electron localization function. The phonon dispersion curves, phonon density of states and thermodynamic properties of $ZrX_2$ (X= H, D, T) compounds were evaluated based on frozen phonon method. Both the Raman and infrared active vibrational modes of $ZrX_2$ at $\Gamma$-point were analyzed which showed significant isotopic effect on $ZrX_2$ compounds. For example, the energy gap between optical and acoustic modes reduces for $ZrT_2$ than $ZrD_2$ and $ZrH_2$. The formation energies of $ZrX_2$ compounds, including the ZPE contributions, were -113.97, -126.50 and -132.04 kJ/(mole of compound) for X = H, D and T, respectively.

## 1. Introduction

The interaction between transition metals, their alloys with hydrogen is an interesting matter of research [1]. The Zr-H system has drawn considerable interest in recent years because of its significance in development of suitable solid state storage materials for hydrogen isotopes and hydride embrittlement of zirconium-based cladding materials in nuclear reactors [2-5]. Specifically, Zr and its alloys like ZrCo are suitable candidate materials for the storage of the heaviest hydrogen isotope (tritium) [6]. Zirconium absorbs hydrogen to a maximum capacity of H/M = 2. However, there are several sub-stoichiometric phases $ZrH_{2-x}$ of this hydride. In the hydrogen absorption-desorption cycle of ZrCo alloy, it has been found that $ZrH_2$ is one of the disproportionated hydride phase which reveal the unusual thermodynamic stability of this compound over the hydrides of its alloys. Similarly, in a nuclear reactor using zirconium alloys as the cladding material, it has been observed that hydrogen/deuterium gets picked up by this material forming the substoichiometric $ZrH_{2-x}$ phase which lead to significant embrittlement and degradation of service life of the clad. Zirconium hydride based fuels such as U-(Th-Np-Am)-Zr-H are also recently being explored because of their excellent stability under irradiation conditions [7]. Hence, it is essential to understand the fundamental thermo-physical properties of $ZrX_2$ (X = H, D, T) for the above applications.

A large number of experimental and theoretical investigations on the Zr-H system have been reported in literature. Weaver et al. [8] have carried out the photoelectron spectroscopy and synchrotron radiation study of $ZrH_x$ (1.63 ≤ x ≤ 1.94) system to understand its electronic structure. Zuzek et al. [9] have constructed the Zr-H phase diagram. At ambinet temperature and pressure, zirconium hydride crystallizes in a fct structure with space group I4/mmm (No.139) whereas the fcc fluorite type structure with space group Fm3m (No.225) is considered as the metastable phase. The phase transition of $ZrH_2$ from cubic to tetragonal structure has been theoretically investigated by Switendick [10]. Recently, several researchers have studied the total energy and electronic structure accurately for Zr-H system using both LDA and GGA exchange correlation functional [11,12]. Weiyi et al. [13] reported the structural and thermodynamic properties of $ZrH_2$ using FP-LAPW method and Debye-Grüneisen quasi-harmonic model. The ground state properties of fcc, fct (c/a = 1.111) and fct (c/a = 0.885) $ZrH_2$ was studied by Zhang et al. [14]

using the first principles method with GGA. The mechanical and structural stability of $ZrH_2$ were investigated by Wang et al. [15] using the plane-wave based pseudopotential method under the framework of density functional theory. Yamanaka *et al.* [16] have reported the electronic, mechanical, electrical and thermal characteristics of zirconium hydride and deuteride. Flotow and Osborne [17] have experimentally measured the heat capacity of $ZrH_2$ and $ZrD_2$ from 5 to $350^oK$ and calculated the hydrogen vibration frequency in $ZrH_2$. To the best of our knowledge, a comparative investigation on $ZrX_2$ (X = H, D, T) compounds, highlighting the effect of isotopes on the vibrational and thermodynamic properties, is not available. Moreover, we are unaware of any study that has reported the physico-chemical properties of the $ZrT_2$ compound. Motivated by these objectives here we report the structural, electronic, vibrational and thermodynamic properties of $ZrX_2$ (X = H, D and T) using '*state of the art*' first principles method which will enable us an understanding of the isotopic effect on the thermo-physical properties of these compounds.

## 2. Calculation methods

All the present calculations were performed using the plane wave-pseudopotential method under the framework of density functional theory as implemented in the Vienna *ab-initio* simulation package (VASP) [18-20]. The electron-ion interaction and the exchange correlation energy were described under the projector-augmented wave (PAW) [21,22] method and the generalized gradient approximation (GGA) of Perdew-Burke-Ernzerhof (PBE) [23], respectively. The valence electronic configuration of Zr and H were set to $5s^14d^3$ and $1s^1$, respectively. The energy cut off for the plane wave basis set was fixed to 500 eV. The ionic optimization was carried out using the conjugate gradient scheme and the forces on each ion was minimized upto 5meV/Å [24,25]. The k-point sampling in the Brillouin zone (BZ) was treated with the Monhorst-Pack scheme [26], using a 6x6x6 k-mesh. Total energies of each relaxed structure using the linear tetrahedron method with Blöchl corrections were subsequently calculated in order to eliminate any broadening-related uncertainty in the energies [27].

To begin with the lattice-dynamical calculations, the lattice parameters of $ZrH_2$ have been optimized using VASP code. In the phonon calculations, the optimized structures of $ZrH_2$ have been used. The phonon frequencies of $ZrH_2$ and its isotopic analogues are calculated by the PHONON program using the forces based on the VASP package. The PHONON package calculates phonon frequencies by supercell calculation of forces on the atoms [28]. A 3x3x2 supercell of ZrH$_2$ containing total 108 atoms have been used for the phonon calculations. A small displacement of 0.02 Å have been given to the atoms present in the supercell of ZrH$_2$. The phonon dispersion curve, phonon density of states and temperature dependent thermodynamic functions of the compounds under study are obtained by using the calculated phonon frequencies.

## 3. Results and discussion

### *3.1 Structural properties*

The crystal structure of α-Zr is hexagonal closed pack (hcp) with space group *P63/mmc*. The hydrogenation of α-Zr produces different hydrides like ζ-Zr$_2$H, γ-ZrH, δ-ZrH$_{1.5}$ and ε-ZrH$_2$ at different conditions. As reported by Flotow *et. al* [17], at 25 $^0$C the ZrH$_2$ compound exist both as face-centered tetragonal (fct) and body centered tetragonal (bct) crystal structure. We have optimized the bct structure with lattice constants $a$ = 3.518 Å and $b$ = 4.447 Å as shown in fig. 1. In the unit cell the Zr and H atoms are placed in the Wyckoff position 2a (0, 0, 0) and 4d (0, 0.5, 0.25), respectively. In this structure (Fig. 1) each Zr atom is surrounded by eight H atoms forming a tetragonal moiety and each H connects with four Zr atoms to build a tetrahedron. To obtain the ground state structural parameters, the total energy calculation were carried out by varying the cell volume as well as the lattice parameters independently. Based on these energies, the energy vs. volume plot (*E-V*) of the ZrH$_2$ is shown in Fig. 2. The parabolic *E-V* graph is fitted with the Murnaghan equation of state [29] to determine the equilibrium lattice constants and bulk moduli ($B_0$). The bulk modulus of the ZrH$_2$ is found to be 76 GPa which is lower than that of α-Zr (85 GPa). This suggests that hydrogenation of α-Zr reduces the fracture strength of α-Zr. The optimized lattice parameters along with their experimental values for α-Zr and ZrH$_2$ are summarized in table 1. The calculated values are found to be within ±1% of the corresponding experimental values [30-33].

### *3.2 Energetics and electronic properties*

The formation of ZrH$_2$ from its constituent elements is described by the chemical reaction:

Zr(s) + H$_2$(g) = ZrH$_2$(s) 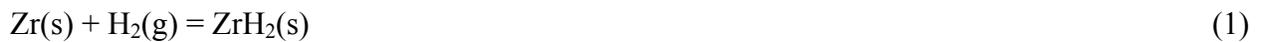 (1)

Hence, the enthalpy of formation of ZrH$_2$ from the elements at any temperature (*T*) can be calculated using the following equation:

$$\Delta_f H (ZrH_2, T) = E_{tot} (ZrH_2, T) - E_{tot} (Zr, T) - E_{tot} (H_2, T) \qquad (2)$$

The total energy of the reactants and products are calculated in this study at $T = 0$ K using the VASP code described above. Using the calculated total energies of Zr, $H_2$ and $ZrH_2$ in equation (2), the formation energy of $ZrH_2$ at 0 K is found to be -159.79 kJ/mol without ZPE contribution. The high negative value of formation energy of $ZrH_2$ compared to Zr depicts its high thermodynamic stability. The formation energy at 0 K is important as this data in combination with the temperature dependent heat capacity data of reactants and products in reaction (1) can be used to derive the temperature dependent enthalpy of hydrogenation reaction of zirconium. Further, the derived values of temperature dependent enthalpy and entropy can be used to calculate the equilibrium dissociation pressure of $ZrH_2$ using the Van't Hoff equation.

In order to describe the nature of chemical bonding in the $ZrH_2$, the orbital projected density of states (DOS) of $ZrH_2$ have been calculated and shown in Fig. 3 which depicts that $4d$-orbital of Zr and $1s$-orbital of hydrogen are majorly participating in the bonding of $ZrH_2$. Furthermore, $1s$-orbital spectrum of hydrogen is more localized (mainly ranging from -3 to -9 eV) compared to the $d$-orbital spectrum of Zr, which is more delocalized over a wide range from -8 to 10 eV. It is seen that at the Fermi level of $ZrH_2$, the $d$-orbital contribution of Zr is dominant whereas the $s$-orbital of hydrogen is contributing majorly below the Fermi energy. The presence of finite DOS at Fermi level in $ZrH_2$ signifies that it is retaining the metallic nature even after the hydrogenation of metallic zirconium. The calculated total DOS of $ZrH_2$ at the Fermi level is $N(E_f) = 0.0813$ states per eV-f.u. The coefficient of electronic specific heat is calculated as $\gamma = (\pi^2/3) \cdot k_B^2 \cdot N(E_f)$, where $k_B$ is the Boltzmann's constant. Using this relation $\gamma$ value of $ZrH_2$ is found to be 0.0317mJ/mol-$K^2$. The understanding of chemical bonding in $ZrH_2$ have been further described by the charge density distribution and electron localization function (ELF) as shown in Fig. 4 and Fig. 5. The color bar graph shows relative electron density across the figures. In Fig. 4, it is seen that more electronic charges are localized at H sites which have been donated by Zr atoms resulting the hydride to be metallic in nature. A different perspective on bond character can be depicted by the ELF, which has been described to be very useful in revealing atomic shell structure and bond charge in crystal systems [34-36]. The ELF is a position dependent function the values of which vary within the range $0 \leq ELF \leq 1$. The ELF value of 0.5 signifies the electron-gas like pair probability. In Fig. 5, the ELF value reaches its highest values of ~0.9 (red) in the vicinity of 1s core state of H atoms. The ELF values decrease in the order ~0.9

(red)<~0.75 (yellow) < ~0.65 (green) < ~0.3 (sky) < ~0.05 (blue) as we move from H sites to Zr sites. The ELF of ZrH$_2$ shows that there is significant charge transfer from Zr to H atoms indicating more ionic bond character which is also consistent with the charge density contour given in Fig. 4.

## *3.3 Vibrational properties and phonon dispersion curves*

According to lattice vibration theory [37], vibrational frequency $\omega$ is a function of both direction and magnitude of wave vector $q$, which is described by the dispersion relation:

$$\omega = \omega_j(q) \tag{3}$$

The subscript j is the branch index. A crystal lattice with n atoms per unit cell has 3n branches, three of which are acoustic modes and the remainders are optical modes. The dispersion curve exhibits symmetry properties in $q$-space, which enables us to restrict consideration to the first Brillouin zone only. The lattice vibration mode with $q \approx 0$ plays a dominant role for Raman scattering and infrared absorption [38]. For this reason, the vibrational frequency with $q = 0$, *i.e.* at the centre $\Gamma$ point of the first Brillouin zone, is called as normal mode of vibration. Since the tetragonal structure of ZrX$_2$ (X= H, D, T) contains 3 atoms per unit cell, so there are nine normal modes of vibrations, which includes three low frequency acoustic modes and six high frequency optical modes. The light atom H has larger displacement amplitude which corresponds to high frequency optical modes and the slow moving heavy atom Zr corresponds to low frequency optical modes. According to group theory [39], the irreducible representations of normal modes at Brillouin zone centre ($\Gamma$ point) can expressed as

2a: $\Gamma_{aco} = 2E_u(IR) + A_{2u}(IR)$ (for Zr)

4d: $\Gamma_{opt} = 2E_u(IR) + A_{2u}(IR) + 2E_g(R) + B_{1g}(R)$ (for H)

where (IR) and (R) stands for infra red active and Raman active modes respectively; subscript *u* and *g* represents antisymmetric and symmetric modes respectively with respect to the center of inversion. The calculated phonon frequencies at the $\Gamma$ point of ZrH$_2$ are listed in table 2.

The phonon dispersion curves show how the phonon energy depends on the q-vectors along the high symmetry directions in the Brillouin zone. This information can be obtained experimentally from neutron scattering experiments on single crystals. The phonon dispersion curves and corresponding phonon density of states (PHDOS) at 0 K for ZrX$_2$ (X= H, D, T) have been obtained by plotting vibrational frequencies along the high symmetry directions, which are shown in Figs. 6-8. It is shown that the dispersion curves and corresponding PHDOS resemble each other for $ZrH_2$, $ZrD_2$ and $ZrT_2$ compounds. Though the acoustic frequencies for $ZrX_2$ (X= H, D, T) are negative but their values are very small in magnitude, say, close to zero and optical phonon frequencies are positive. This indicated that the tetragonal structure of $ZrX_2$(X= H, D, T) is dynamically stable. Despite some differences in the exact position of the optical modes among the $ZrX_2$ (X = H, D, T) compounds, the spectral feature of the phonon dispersion curves and PHDOS show similar pattern. It is known that the mass difference between the constituent elements of the compound strongly affects the maximum and minimum values of the acoustic and optical branches. Thus a frequency gap between these two branches is formed, as can be seen in Figs. 6-8. It is observed that the gap between the high frequency optical modes and low frequency acoustic modes decreases as heavier isotopes are substituted in $ZrX_2$. The range of the frequency gap varies from ~24 THz ($ZrH_2$) to ~15 THz ($ZrD_2$) to ~11 THz ($ZrT_2$). However, the PHDOS frequency of Zr for all the three compounds remains constant in the range ~0 to 7 THz.

### 3.4 Thermodynamic properties

The temperature-dependent thermodynamic functions of a crystal, such as the internal energy (*E*), entropy (*S*), Helmholtz free energy (*F*) and constant volume heat capacity ($C_V$), is calculated from their phonon density of states as a function of frequencies using the following formulas:

$$F = 3nNk_BT \int_0^{\omega_{max}} \ln\left\{2\sinh\frac{\hbar\omega}{2k_BT}\right\} g(\omega)d\omega \tag{4}$$

$$E = 3nN\frac{\hbar}{2}\int_0^{\omega_{max}} \omega \coth\left(\frac{\hbar\omega}{2k_BT}\right) g(\omega)d\omega \tag{5}$$

$$S = 3nNk_B \int_0^{\omega_{max}} \left[\frac{\hbar\omega}{2k_BT}\coth\frac{\hbar\omega}{2k_BT} - \ln\left\{2\sinh\frac{\hbar\omega}{2k_BT}\right\}\right] g(\omega)d\omega \tag{6}$$

$$C_V = 3nNk_B \int_0^{\omega_{max}} \left(\frac{\hbar\omega}{2k_BT}\right)^2 \csc h^2\left(\frac{\hbar\omega}{2k_BT}\right) g(\omega)d\omega \tag{7}$$

The variation of *F*, *E*, *S*, $C_V$ are shown in Fig. 9 (a-d) upto 800 K, which is below the decomposition temperature of $ZrH_2$. It may be noted that the *F* and the *E* at 0 K represents the zero point energy, which can be calculated from the expression as $F_0 = E_0 = 3nN \int_0^{\omega_{max}} \left(\frac{\hbar\omega}{2}\right) g(\omega)d\omega$, where *n* is the number of atoms per unit cell, *N* is the number of unit cells, $\omega$ is the phonon frequencies, $\omega_{max}$ is the maximum phonon frequency, and g($\omega$) is the normalized phonon density of states with $\int_0^{\omega_{max}} g(\omega)d\omega = 1$.

The calculated zero point energies (ZPE) are 45.08, 33.29 and 27.75 kJ/mol for ZrH$_2$, ZrD$_2$ and ZrT$_2$, respectively. The enthalpy of formation ($\Delta_f H$) at 0 K for ZrH$_2$ is found to be -159.79 kJ/mol without ZPE contribtution. After including the ZPE correction, the heat of formation of ZrH$_2$ changes from -159.79 to -113.97 kJ/mol. Similarly, ZPE corrected $\Delta_f H$ for ZrD$_2$ and ZrT$_2$ are -126.5 and -132.04 kJ/mol, respectively. Sieverts *et. al* [40] and Flotow *et. al* [17] reported the enthalpies of formation of ZrH$_2$ and ZrD$_2$ are -169.65 and -173.84 kJ/mol, respectively at 298 K. However, they did not include the ZPE correction. It is interesting to mention that, although ZrH$_2$ and its isotopic analogues ZrD$_2$ and ZrT$_2$ have the same crystal and electronic structure, ZrT$_2$ and ZrD$_2$ are more stable than ZrH$_2$. This is in line with the prediction made by Hu et al. [41] and Frankcombe et al. [42]. Fig. 9(a) shows that the free energy (*F*) for all three hydrides decreases gradually with increase in temperature. In contrast, the internal energy *E* and entropy *S* increases with increase in temperature as shown in Fig. 9(b) and 9(c). Moreover, a good agreement between the experimental and calculated values of entropy (S) was obtained for ZrH$_2$ and ZrD$_2$. After establishing the accuracy of the calculated entropy for ZrH$_2$ and ZrD$_2$, it can be argued that the calculated entropy of ZrT$_2$, for which experimental data is not available, will be reasonably accurate.

We have also calculated the heat capacities of ZrX$_2$ (X = H, D, and T), which is an interesting parameter to understand the thermodynamic stability of solids. The temperature dependence of heat capacity ($C_V$) of ZrX$_2$ (X= H, D, T) is shown in Fig. 9(d). It is seen that at low temperature, upto 600K, the heat capacities of ZrX$_2$ (X = H, D, and T) increase rapidly with increase in temperature and thereafter increases slowly up to 800 K, and attain the saturation value of ~70 J/mol.K. This is the Dulong-Petit classical limit. From the fig. 9(d) it is seen that upto 100K, the variation of $C_v$ versus *T* curve follows the same trend for all ZrX$_2$ compounds, but above 100K it shows different trend: $C_v$(ZrT$_2$) > $C_v$(ZrD$_2$) > $C_v$(ZrH$_2$). In addition, the heat capacity plot shows a broad hump around 100 K for all three compounds. This is in agreement with the previously reported results of Flotow and Osborne [17]. At low temperature, below 100 K, the acoustic modes in which the Zr and H atoms are vibrating 'in-phase' dominate the heat capacity contribution. At higher temperature, above 100 K, the optical modes of vibration in which the light H atoms vibrate against the heavy and almost stationary Zr atoms, dominates the heat capacity. Hence, around 100 K, this transition from predominant acoustic vibration to optical vibration cause a change in heat capacity function thus showing a broad hump. Now we compare the calculated heat capacity with the available experimental results. As it is not possible to measure the $C_V$ directly from experiments, so the calculated $C_V$ was compared with the experimentally reported $C_P$ values. In this context, it needs to be mentioned that although $C_p$ values are available for $ZrH_2$ and $ZrD_2$, no experimental results are available for $ZrT_2$. For solids, the relation between $C_P$ and $C_V$ is $C_P - C_V = \alpha^2 TVB$, where $\alpha$ is the thermal expansion coefficient, $V$ is the molar volume, and B is the bulk modulus of the system. It is reported that the difference between $C_P$ and $C_V$ is on the order of a few percent of $C_V$ [43]. Since the data of $C_v$ calculated from phonon DOS represents only the vibrational contribution, the calculated $C_v$ data of $ZrH_2$ from 0 - 20 K have been fitted with the equation $C_V = 1943.9 \left(T/\theta_D\right)^3$ where $\theta_D$ is the Debye temperature. The Debye temperature $\theta_D$ obtained by fitting the $C_V$ versus $T$ graph of $ZrH_2$ is 324.4 K, which is in good agreement with the experimentally reported value of 311.4 K for $ZrH_2$ [17]. The calculated $C_V$ and $S$ values of $ZrH_2$ and $ZrD_2$ at 298 K are compared with the experimental data in table 3. Although the calculated $C_V$ is marginally different than the experimental $C_P$ values, but the calculated $S$ values are in excellent agreement with the experimental data. The difference between the calculated and experimental values of heat capacity at high temperature is attributed to two reasons; (i) the anharmonicity effect and (ii) electron phonon coupling.

## 4. Conclusions

In summary, we have investigated the electronic, vibrational, and thermodynamic properties of zirconium hydrides, $ZrX_2$ (X = H, D, T), with a focus to the effect of isotopic substitution. The nature of chemical bonding was analyzed through charge density distribution, electronic density of states, and electron localization function. The results showed significant electronic charge transfer from Zr to the X site, indicating ionic bond character. The phonon frequencies, phonon density of states and phonon dispersion curves were obtained using the direct super-cell method and the Raman and infrared active modes of all the compounds were assigned. Using the phonon frequencies, the thermodynamic functions were determined within the framework of harmonic approximation. A good agreement was found between the experimental results and the calculated values. On the basis of the ZPE corrected enthalpies of formation ($\Delta_f H$ at 0 K) for all three compounds, it is found that $ZrT_2$ and $ZrD_2$ are more stable than $ZrH_2$. Thus the effect atomic mass of different isotopes on the thermodynamic properties was established. Based on the results it is envisaged that the ZPE corrected enthalpy of formation data at 0 K coupled with the high temperature data of $ZrX_2$ (X= H, D, T) will help us to calculate the equilibrium dissociation pressure of $X_2$ (X= H, D, T) for practical applications.

**Acknowledgements**

The authors are thankful to the members of the Computer Division, BARC, for their kind cooperation during this work and Dr. K.L. Ramakumar, Director, Radiochemistry and Isotope Group, for his encouragement.

**References**

[1] Fukai Y, Metal Hydrogen System, Basics Bulk Properties, 2005 Springer, Berlin.

[2] Züttel A, Materials for hydrogen storage, 2003 Mater. Today, **9** 24-33.

[3] Tapping R L, Gendron T S, A survey of in-reactor data relevant to the corrosion and hydriding of zirconium alloys in annulus gas environments; 1988, AECL-RC101 COG-88-136.

[4] Bickel P W, Berlincourt T G, 1970 Phys. Rev. B **2** 4807.

[5] Holliger L, Legris A, Besson R, 2009 Phys. Rev B **80** 094111.

[6] Chattaraj D, Parida S C, Dash S, Majumder C, 2012 Int. J. Hydrogen Energy **37** 18952.

[7] Huang J, Tsuchiya B, Konashi K, Yamawaki M, 2000 J. Nucl. Sci. Techol. **37** 887.

[8] Weaver J H, Peterman D J, Peterson D T, Franciosi A, 1981 Phys. Rev. B **23** 1692.

[9] Zuzek E, Abriata J P, Martin A S, Manchester F D, 1990 Bull. Alloy Phase Diagrams, **11** 385.

[10] Switendick A. C, 1984 J. Less-Common Met. **101** 191.

[11] Wolf W, Herzig P, 2000 J. Phys.: Condens. Matter **12** 4535.

[12] Quijano R, Coss R de, Singh D J, 2009 Phys. Rev. B **80** 184103.

[13] Weiyi R, Pingchuan X, Weiguo S, 2010 Physica B, **405** 2057.

[14] Zhang P, Wang B, He C, Zhang P, 2011 J. Comp. Mat. Sci. **50** 3297.

[15] Wang F, Gong H R, 2012 Int. J Hydrogen Energy **37** 9688.

[16] Yamanaka S, Yamada K, Kurosaki K, Uno M, Takeda K, Anada H, Matsuda T, Kobayashi S, 2002 J. Alloys. Comp. 2002 **330-332** 99-104.

[17] Flotow H E, D.W. Osborne 1961 J. Chem. Phys. **34** 1418.

[18] Kresse G, Hafner J, 2004 Phys Rev B **49** 14251-69.

[9] Kresse G, Furthmüller J, 1996 Comput Mater Sci 6 15-50.

[20] Kohn W, Sham L J, 1965 Phys Rev A **140** 1133-8.

[21] Blöchl P E, 1994 Phys Rev B **50** 17953-79.

[22] Kresse G, Joubert D, 1999 Phys Rev B **59** 1758-75.

[23] Perdew J P, Burke K, 1996 Phys Rev Lett **77** 3865-8.

[24] Feynman R P, 1939 Phys Rev **56** 340-3.

[25] Hellman H, 1937 Introduction to Quantum Chemistry. Leipzig: Deuticke.

[26] Monkhorst H J, Pack J D, 1976 Phys. Rev. B **13** 5188-5192.

[27] Blöchl P E, Jepsen O, Andersen O K, 1994 Phys Rev B **49** 16223-33.

[28] Parlinski K, Li Z Q, Kawazoe Y, 1997 Phys. Rev. Lett. **78** 4063.

[29] Murnaghan F D, 1944 Proc. of Nat. Acad.of Sc. USA **30** 244.

[30] Niedzwiedz K, Nowak B, Zogal O J, 1993 J. Alloys Comp. **94** 47.

[31] Russel R B, 1953 J. Appl. Phys. **24** 232.
[32] Sikka S K, Vohra Y K, Chidambaram R, 1982 R. Prog. Mater. Sci., **27** 245.
[33] W. Zhu, R. Wang, G. Shu, P. Wu, H. Xiao, J. Phys. Chem. C., 114 (2010) 22361.
[34] Becke A D, Edgecombe K E, 1990 J. Chem. Phys. **92** 5397.

[35] Silvi B, Savin A, 1994 Nature **371** 683.

[36] Savin A, Nesper R, Wengert S, Fässler T F, 1997 Angew. Chem. Int. Ed. Engl. **36** 1808.

[37] Omar M A, 1975 Elementary Solid State Physics: Principles and Applications, Addision-Wesley Publishing Company, Reading, Massachusetts.

[38] Zhang G Y, Lan G X, 1991 Lattice Vibration Spectroscopy, High Education Press, Beijing.

[39] Hahn T, 1983 The International Table for Crystallography, Vol. A: Space Group Symmetry, Reidel, Dordrecht. .

[40] Siverts A, Gotta A, Halberstadt S, 1930 Z. anorg. u. allgem. Chem. **187** 159.

[41] Hu C H, Chen D M, Wang Y M, Yang K, 2008 J. Alloys Comp. **450** 369.

[42] Frankcombe T J, Kroes G J, 2006 Chem. Phys. Lett. **423** 102.

[43] Ke X Z, Tanaka I, 2005 Phys. Rev. B **71** 024117.

[44] Bale C W, Chartrand P, Decterov S A, Eriksson G, Hack K, Ben Mahfoud R, Melançon J, Pelton A D and Petersen S, 1976-2012 "FactSage Thermochemical Software and Databases", version 6.3.

**Figure Captions**

Fig.1 Crystal structure of $ZrH_2$ (big spheres for Zr and small spheres for H)

Fig. 2 Variation in energy as a function of cell volume of $ZrH_2$.

Fig. 3 The orbital projected density of states (DOS) of $ZrH_2$.

Fig. 4 The charge density contour of $ZrH_2$ along the (100) plane.

Fig. 5 The electron localization function of $ZrH_2$ along the (100) plane.

Fig. 6 Calculated phonon dispersion curves and total density of states for $ZrH_2$ along the main symmetry directions in BZ.

Fig. 7 Calculated phonon dispersion curves and total density of states for $ZrD_2$ along the main symmetry directions in BZ.

Fig. 8 Calculated phonon dispersion curves and total density of states for $ZrT_2$ along the main symmetry directions in BZ.

Fig. 9 The calculated phonon contribution to the Helmholtz free energy $F$ (a), the phonon contribution to the internal energy $E$ (b), the entropy $S$ (c), and the constant-volume specific heat $C_v$ (d) for $ZrH_2$, $ZrD_2$ and $ZrT_2$.

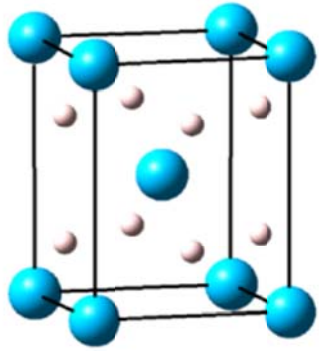

**Fig. 1**

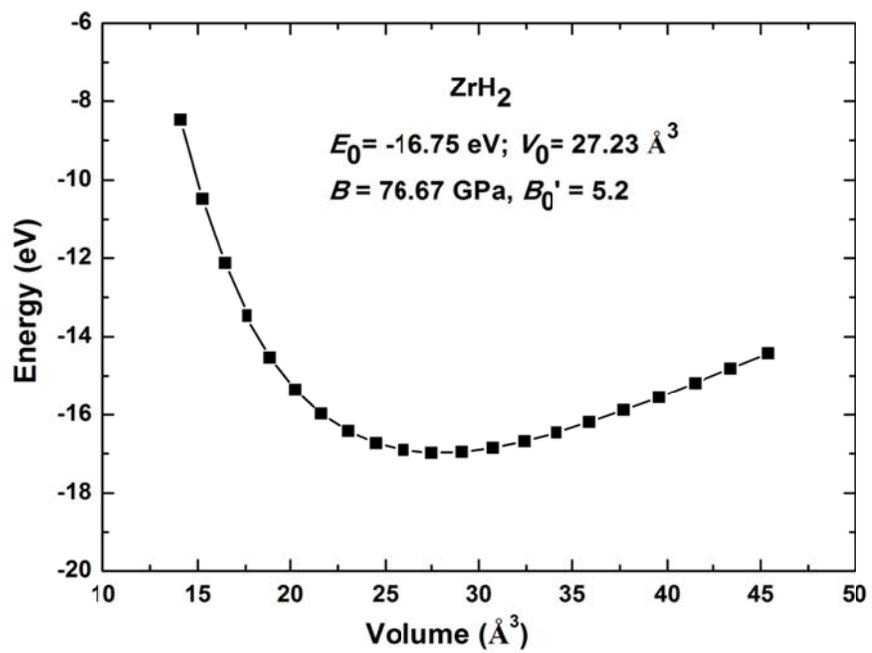

**Fig. 2**

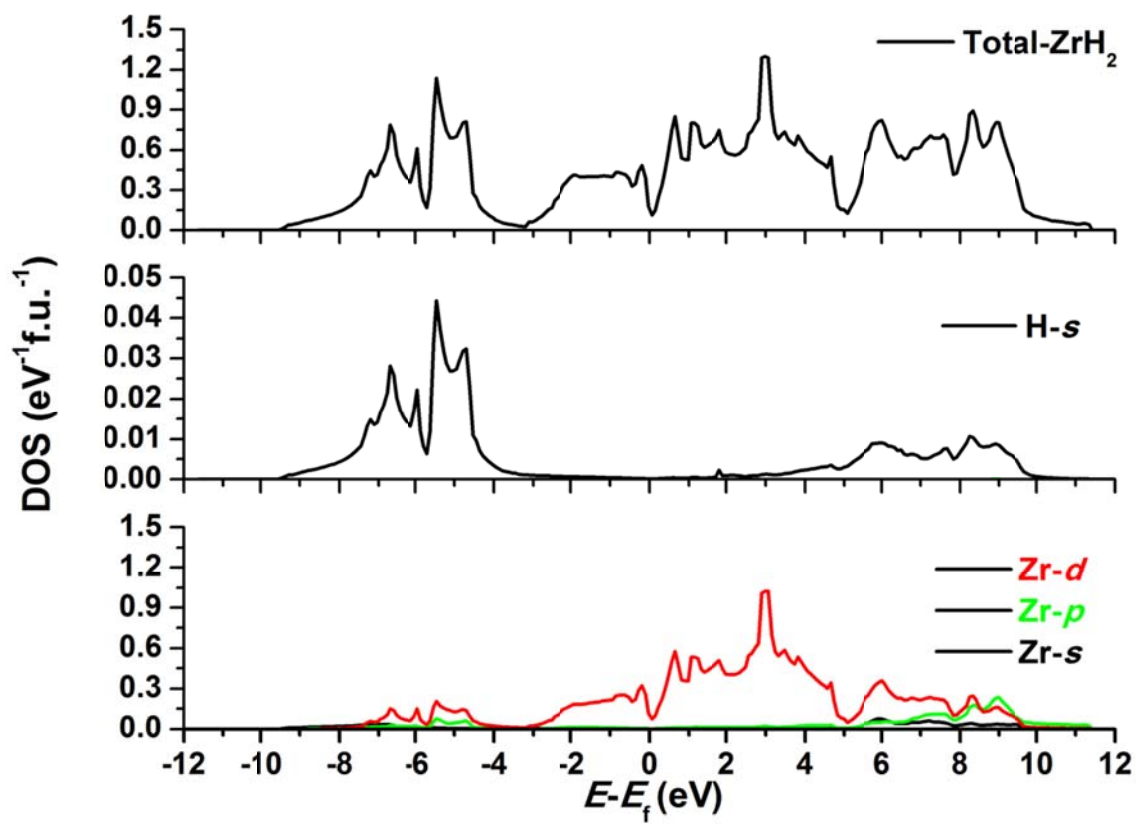

**Fig. 3**

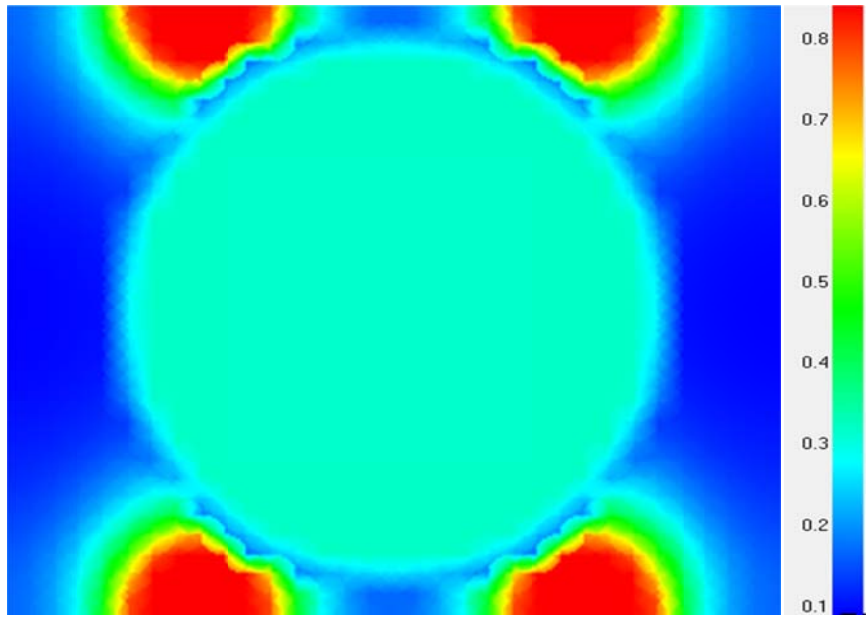

**Fig. 4**

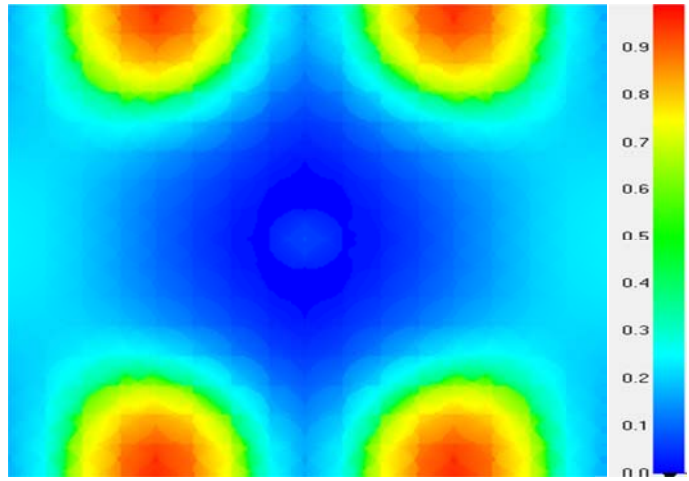

**Fig. 5**

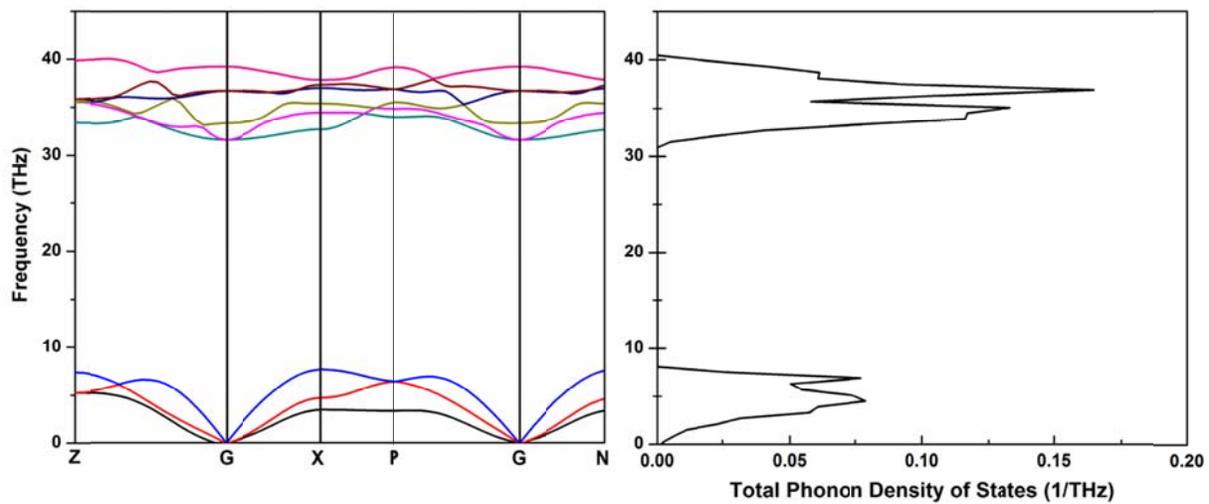

**Fig. 6**

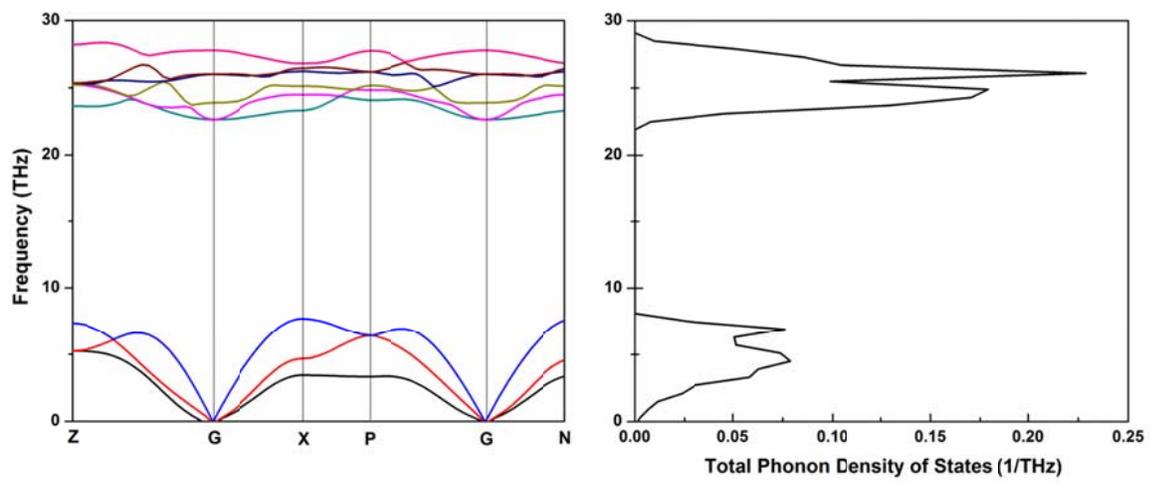

**Fig. 7**

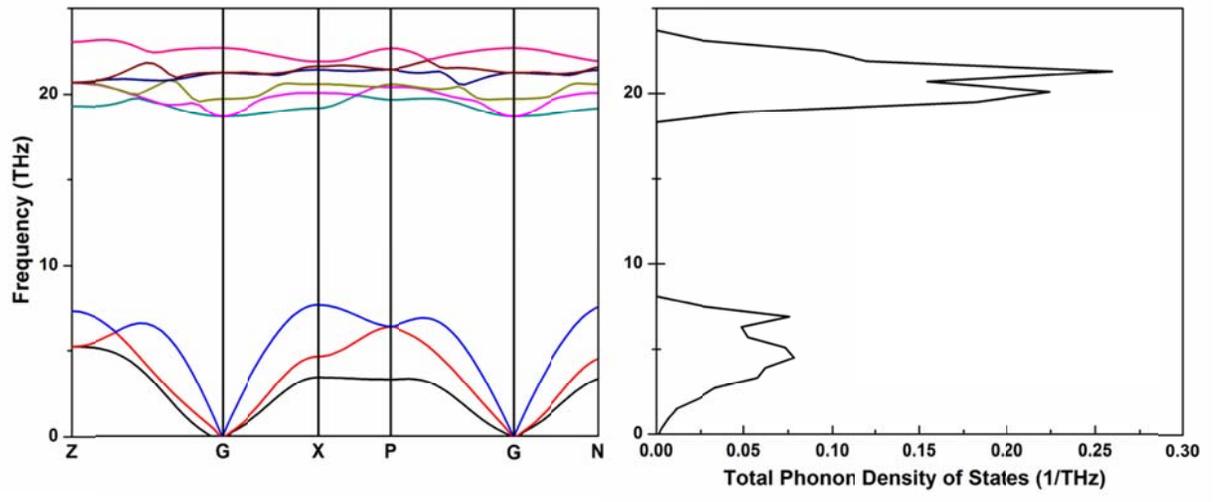

**Fig. 8**

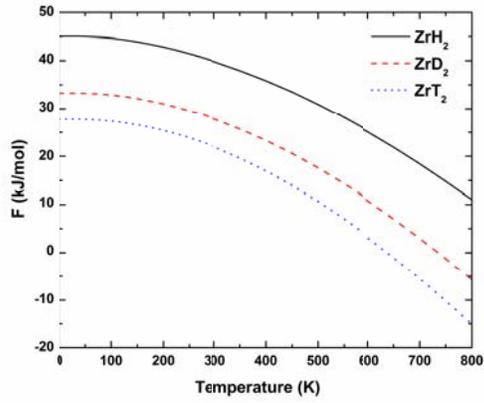
(a)

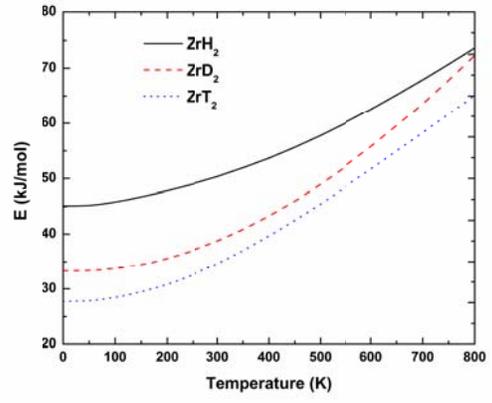
(b)

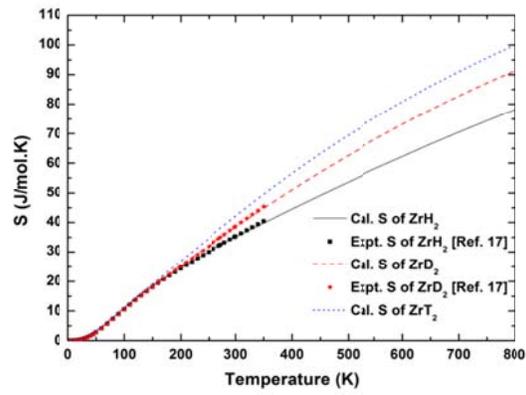
(c)

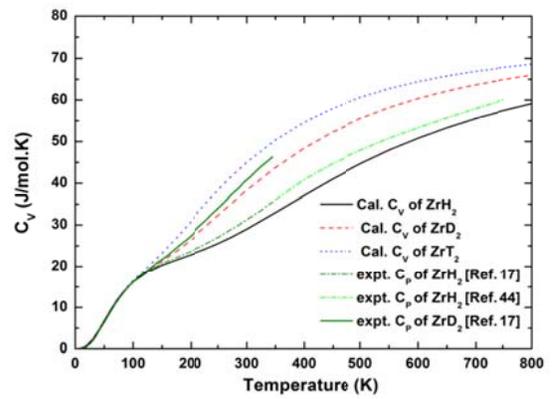
(d)

**Fig. 9**

**Table Captions**

Table 1: Optimized lattice constants along with the available experimental values (in Ǻ) for α-Zr and ZrH$_2$.

Table 2: Phonon frequency at the $\Gamma$ point of ZrH$_2$.

Table 3: The calculated and experimental specific heat and entropy of ZrH$_2$, ZrD$_2$ and ZrT$_2$.

| System | this work (at 0 K) | | Expt. (at 298 K) | | previous work (at 0 K) | |
|---|---|---|---|---|---|---|
| | *a* | *c* | *a* | *c* | *a* | *c* |
| α- Zr | 3.230 | 5.169 | 3.233[a] | 5.150[a] | 3.223[c] | 5.157[c] |
| | | | 3.231[b] | 5.147[b] | | |
| ZrH$_2$ | 3.538 | 4.442 | 3.518[d] | 4.447[d] | - | - |

[a] Reference [31].
[b] Reference [32].
[c] Reference [33].
[d] Reference [30].

Table 1

| Branch | | Wyckoff | Mode | Multiplicity | Frequency (THz) | Active mode |
|---|---|---|---|---|---|---|
| ZrH$_2$ | Acoustic | 2a (Zr) | $E_u$ | 2 | -0.343 | Infrared |
| | | | $A_{2u}$ | 1 | -0.121 | Infrared |
| | Optical | 4d (H) | $E_u$ | 2 | 31.866 | Infrared |
| | | | $A_{2u}$ | 1 | 33.645 | Infrared |
| | | | $E_g$ | 2 | 36.744 | Raman |
| | | | $B_{1g}$ | 1 | 39.364 | Raman |
| ZrD$_2$ | Acoustic | 2a (Zr) | $E_u$ | 2 | -0.340 | Infrared |
| | | | $A_{2u}$ | 1 | -0.120 | Infrared |
| | Optical | 4d (D) | $E_u$ | 2 | 22.774 | Infrared |
| | | | $A_{2u}$ | 1 | 24.049 | Infrared |
| | | | $E_g$ | 2 | 25.982 | Raman |
| | | | $B_{1g}$ | 1 | 27.834 | Raman |
| ZrT$_2$ | Acoustic | 2a (Zr) | $E_u$ | 2 | -0.336 | Infrared |
| | | | $A_{2u}$ | 1 | -0.119 | Infrared |
| | Optical | 4d (T) | $E_u$ | 2 | 18.790 | Infrared |
| | | | $A_{2u}$ | 1 | 19.845 | Infrared |
| | | | $E_g$ | 2 | 21.214 | Raman |
| | | | $B_{1g}$ | 1 | 22.727 | Raman |

Table 2

| System | Calculated (J mol$^{-1}$K$^{-1}$) | | Experimental (J mol$^{-1}$K$^{-1}$) | |
| --- | --- | --- | --- | --- |
| | $C_V$ (at 298 K) | $S$ (at 298 K) | $C_P$ (at 298.15 K) | $S$ (at 298.15 K) |
| ZrH$_2$ | 28.88 | 34.87 | 30.98 ± 0.06[a], 30.96[b] | 34.83±0.08[a], 35.02[b] |
| ZrD$_2$ | 38.10 | 38.27 | 40.34±0.08[a] | 38.40±0.08[a] |
| ZrT$_2$ | 44.76 | 41.93 | - | - |

[a]Ref. [17]
[b]Ref. [44]

Table 3